\documentclass[aps,prl,twocolumn,superscriptaddress]{revtex4-1}
\usepackage{amsmath,amssymb,graphicx,natbib,aas_macros,bm}
\usepackage{enumerate}

\newcommand{\keV}{\ensuremath{\mathrm{keV}}}
\newcommand{\MeV}{\ensuremath{\mathrm{MeV}}}
\newcommand{\GeV}{\ensuremath{\mathrm{GeV}}}

\begin{document} 

\title{Non-Abelian Discrete Dark Matter}

\author{Adisorn Adulpravitchai}
\email{a.adulpravitchai@mpi-hd.mpg.de}
\affiliation{Max-Planck-Institut f${\ddot{\it u}}$r Kernphysik, Postfach 10 39 80, 69029 Heidelberg, Germany}

\author{Brian Batell}
\email{bbatell@perimeterinstitute.ca}
\affiliation{Perimeter Institute for Theoretical Physics, Waterloo,
ON, N2L 2Y5, Canada}

\author{Josef Pradler}
\email{jpradler@perimeterinstitute.ca}
\affiliation{Perimeter Institute for Theoretical Physics, Waterloo,
ON, N2L 2Y5, Canada}

\begin{abstract}
  We consider the minimal model in which dark matter is stabilized by
  a non-Abelian discrete symmetry. The symmetry group is taken to be
  $D_3\cong S_3$, which is the smallest non-Abelian finite group.  The
  minimal model contains (nontrivial) singlet and doublet scalar
  representations of $D_3$ which couple to the Standard Model fields
  via the Higgs portal. This construction predicts two species of dark
  matter over much of the parameter space. Nontrivial interactions
  under $D_3$ lead to a novel thermal history of dark matter, while
  the multi-component nature of dark matter can be tested by future
  direct detection experiments.
\end{abstract}

\maketitle

Understanding the nature of the cosmological dark matter (DM) that
constitutes one quarter of the energy density of the universe is a
central goal of particle physics today~\cite{review}. While there is
little room left to doubt the existence of DM, its microscopic
properties are virtually unknown.  One of the few properties in which
we can be confident is that DM should be stable on time scales greater
than the age of the universe, suggesting the existence of a new
``dark'' symmetry.  But precisely what symmetry stabilizes DM is a
mystery.

Many models employ a discrete $Z_2$ symmetry to stabilize DM.  This
$Z_2$ symmetry is often motivated by the need to suppress dangerous
operators in new physics scenarios that solve the hierarchy problem.
However, given that we have no experimental indication of what new
physics, if any, addresses the naturalness issues in the SM, one may
take a more general perspective regarding DM and the symmetries
responsible for its stability. More pragmatically, the exploration of
alternative stabilizing symmetries is warranted by the prospect of
novel phenomena associated with DM, as such symmetries may predict new
states and interactions.

Indeed, there are many possibilities other than a $Z_2$-parity that
can protect DM against decay. In particular, besides the Abelian
cyclic symmetry $Z_N$~\cite{DGS}, DM may well be stabilized by a non-Abelian
discrete symmetry.  Non-Abelian finite groups have received some
limited attention within the context of DM. Motivated by improved
gauge coupling unification, Ref.~\cite{lisanti} considered an
additional Higgs doublet in a non-Abelian discrete multiplet serving
as DM.  Non-Abelian discrete symmetries also lead to distinct decay
patterns in decaying dark matter scenarios~\cite{Haba:2010ag,
  Kajiyama:2010sb}. \textit{Continuous} non-Abelian symmetries
originating from broken or confined gauge theories can also ensure DM
stability~\cite{Hambye}.  Models of DM stabilized by \textit{Abelian}
discrete symmetries that descend from higher non-Abelian ones have
been motivated by astrophysical anomalies~\cite{cline}, discrete gauge
symmetries~\cite{walker}, and neutrino
physics~\cite{Hirsch:2010ru}. Indeed,
non-Abelian flavor symmetries are
widely used to explain the neutrino oscillation
data~\cite{Schwetz:2008er} (for recent reviews see Refs.~\cite{Altarelli:2010gt,Ishimori:2010au}).
Such non-Abelian discrete symmetries can come from the breaking of
continuous flavor symmetries~\cite{ConSym}
or from orbifold compactification of
extra-dimensions~\cite{OrbifoldSym}.

In this paper we construct the minimal model of DM in which stability
is a consequence of a non-Abelian discrete symmetry. The model follows
in spirit the canonical scalar Standard Model (SM) singlet models of
Refs.~\cite{singlet1,singlet2,singlet3}. The symmetry group we
consider is $D_3$, the dihedral group of order~6, which is the
smallest non-Abelian finite group.  It is isomorphic to $S_3$, the
permutation symmetry of three objects. We determine the minimal field
content of the model and couple it to the SM via renormalizable Higgs
portal interactions.  Over much of the parameter space, the $D_3$
symmetry predicts two species of DM, which can contribute to the
cosmological abundance.  We investigate in detail the cosmology of the
model, finding that nontrivial interactions under $D_3$ in the dark
sector lead to a novel thermal history.  We also analyze the
constraints and prospects at direct detection experiments. In
particular we find that the two-component nature of DM can be
discovered in future ton-scale experiments.

\paragraph{ \bf  \boldmath$D_3$  stabilization of dark matter.}

We start by constructing the minimal model of DM stabilized by a
non-Abelian discrete symmetry. The model is based on the group $D_3$,
which is the smallest non-Abelian discrete symmetry group and
describes the symmetry properties of the equilateral triangle. The
group contains two generators, $A$ and $B$, obeying the properties
\begin{equation}
A^3 =1, \quad B^2 =1, \quad ABA=B,
\end{equation}
and the six group elements are constructed through combinations of the
two generators: $1$, $A$, $A^2$, $B$, $BA$, $BA^2$. The group $D_3$
has two singlet and one doublet representation, denoted, ${\bf
  \underline{1}}_{ \bf 1 }$, ${\bf \underline{1}}_{ \bf 2}$, and ${\bf
  \underline{2}}$ respectively. The singlet ${\bf \underline{1}}_{ \bf
  1 }$ transforms trivially under $D_3$, $A=B=1$, while the generators
of the (nontrivial) singlet representation ${\bf \underline{1}}_{ \bf
  2}$ are $A=1$ and $B=-1$. The generators for the doublet
representation ${\bf \underline{2}}$ are
\begin{eqnarray} 
A= \left( 
\begin{array}{cc}
e^{2\pi i/3} & 0 \\
 0 & e^{- 2\pi i /3 } 
\end{array}
 \right),  ~& 
~ B= \left( 
\begin{array}{cc}
0 & 1 \\
 1 & 0 
\end{array}
 \right).  
\end{eqnarray}
The multiplication rules for the singlet representations are given by
\begin{equation}
 {\bf \underline{1}}_{ \bf 1} \otimes {\bf \underline{1}}_{ \bf 1} = {\bf \underline{1}}_{ \bf 1}, \;\; {\bf \underline{1}}_{ \bf 2} \otimes {\bf \underline{1}}_{ \bf 2} = {\bf \underline{1}}_{ \bf 1}, \; \mbox{and} \; \; {\bf \underline{1}}_{ \bf 1} \otimes {\bf \underline{1}}_{ \bf 2} = {\bf \underline{1}}_{ \bf 2} \;,
\label{singlet}
\end{equation}
whereas the product of two doublets, ${\bf \underline{2}} \otimes {\bf
  \underline{2}}$ decomposes into two singlets and one doublet as
\begin{eqnarray}
& (a_1 b_2 + a_2 b_2) \sim {{\bf \underline{1}}_{ \bf 1}} ,\;\;  (a_1 b_2 - a_2 b_2) \sim {{\bf \underline{1}}_{ \bf 2}} ,\;\; \begin{pmatrix} a_2 b_2 \\ a_1 b_1 \end{pmatrix} \sim {\bf \underline{2}}, & \nonumber \\
&&  
\label{doublet}
\end{eqnarray}
where $(a_1,a_2)^T, (b_1,b_2)^T \sim {\bf \underline{2}}$; for further
formulae of dihedral groups, see, \textit{e.g.},~\cite{Blum:2007jz,
  Frampton:1994rk,Ishimori:2010au}.

The minimal model of DM having nontrivial properties under $D_3$
contains two scalar fields,
\begin{equation}
\eta ,  \qquad 
X \equiv \left( \begin{array}{c}
\chi \\
\chi^*
\end{array}
 \right),
 \end{equation}
 which transform as 
 a singlet ${\bf \underline{1}}_{ \bf 2}$ and doublet ${\bf
   \underline{2}}$, respectively.  The scalar $\eta$ is real while
 $\chi$ is complex, which amounts to a total of three new degrees of
 freedom. Along with the SM Higgs field $H$, the scalar sector of the
 Lagrangian contains kinetic terms and a potential. The general
 renormalizable scalar potential invariant under $D_3$ and the SM
 gauge symmetries may be constructed using
 Eqs.~(\ref{singlet},\ref{doublet}) and reads
\begin{align}
  V & = m_1^2 H^\dag H + \frac{1}{2} m_2^2 \eta^2 + m_3^2 \chi^* \chi
  + \frac{\mu_1}{3!} (\chi^3 + {\chi^*}^3 ) \nonumber \\ &
  + \lambda_1 (H^\dag H)^2 + \frac{\lambda_2}{4} \eta^4 +\lambda_3 (\chi^* \chi)^2 \nonumber \\
  & +\alpha_1 (H^\dag H )\eta^2 + 2 \alpha_2 (H^\dag H)( \chi^* \chi)
  +\alpha_3 \eta^2(\chi^* \chi) \nonumber \\ & + \frac{i\alpha_4 }{3!}
  \eta (\chi^3 -{\chi^*}^3),
\label{potential}
\end{align}
where all parameters in the Lagrangian are real.  We note that that
the theory is invariant under $P$ and $C$---a fact that will simplify
our considerations regarding the relic abundance.

A model with only one scalar field $\eta$ ($\chi$) can provide a
viable theory of DM, but such a theory is equivalent to a theory based
on an Abelian discrete $Z_2$ ($Z_3$) symmetry.  The minimal theory
based on the group $D_3$ contains both scalars, $\eta$ and $\chi$, and
the nontrivial interaction predicted by this symmetry is the last term
in the potential (\ref{potential}), with coupling $\alpha_4$.

We will be interested in the case in which the electroweak symmetry is
spontaneously broken while the $D_3$ discrete symmetry is unbroken,
\begin{equation}
 \langle H \rangle= \frac{1}{\sqrt{2}}\left(
\begin{array}{c}
 0 \\
v 
\end{array}
  \right), \qquad 
 \langle \eta \rangle= 0, \qquad
 \langle \chi \rangle= 0,
\label{vac}
\end{equation}
where $v^2 \equiv -m_1^2/\lambda_1 \simeq (246\,\GeV)^2$ is the
(squared) Higgs vacuum expectation value.  We demand that the
potential is bounded from below, and that the electroweak vacuum is
the global minimum of the potential. These requirements constrain the
values of the coupling constants in the potential.  Qualitatively, the
quartic couplings $\lambda_{1,2,3}$ must be positive, $\alpha_{1,2,3}$
must be greater than some minimum (negative) value, and the magnitude
of $\alpha_4$ must be smaller than some maximum value. Furthermore,
the mass parameters $m_{2,3}^2$ should not be too large and the
magnitude of the cubic coupling $\mu_1$ is bounded from above.
Throughout our analysis, we have verified numerically that our
parameter choices lead to a stable potential and electroweak
vacuum. One may also investigate the vacuum structure of the theory at
higher energy scales using a renormalization group analysis (as, {\em
  e.g.}, in Ref.~\cite{stability}), but here we will be content to
consider Eq.~(\ref{potential}) a low energy effective theory of DM
defined around the weak scale.

Expanding around the vacuum (\ref{vac}) with $v \rightarrow v+h$, we derive the following potential:
\begin{align}
V & =  \frac{1}{2}m_h^2 h^2 + \frac{1}{2} m_\eta^2 \eta^2 + m_\chi^2 \chi^* \chi \nonumber \\  
& + \lambda_1 v h^3 + \alpha_1 v h \eta^2 + 2 \alpha_2 v h (\chi^* \chi) + \frac{\mu_1}{3!} (\chi^3 + {\chi^*}^3 ) \nonumber \\
& + \frac{\lambda_1}{4} h^4 + \frac{\lambda_2}{4} \eta^4 +\lambda_3 (\chi^* \chi)^2 \nonumber \\ 
& + \frac{\alpha_1}{2} h^2\eta^2 + \alpha_2 h^2( \chi^* \chi) 
+\alpha_3 \eta^2 (\chi^* \chi)   \nonumber \\ & +
\frac{i \alpha_4}{3!} \eta (\chi^3 -{\chi^*}^3) ,
\label{Lag2}
\end{align}
where the physical masses are 
\begin{eqnarray}
m_h^2 & \equiv & 2 \lambda_1 v^2, \nonumber \\
m_\eta^2 & \equiv & m_2^2+\alpha_1 v^2, \nonumber \\
m_\chi^2 & \equiv & m_3^2+\alpha_2 v^2. 
\label{physicalmass}
\end{eqnarray}

The nontrivial interaction predicted by the $D_3$ symmetry, with
coupling $\alpha_4$, allows the scalar $\eta$ to decay via $\eta
\rightarrow 3 \chi, 3\chi^*$ if $m_\eta > 3 m_\chi$.  In this case,
$\chi$ is the only stable DM candidate.  If, however, $m_\eta < 3
m_\chi$ these decays are kinematically forbidden, and both $\eta$ and
$\chi$ are stable.  Therefore, the $D_3$ model predicts two species of
DM for $m_\eta < 3 m_\chi$.

\paragraph{\bf Cosmology.}

We now examine the thermal history of the $\eta$ and $\chi$ particles
in the minimal $D_3$ model. The cosmic relic abundances of these
particles are governed by Boltzmann equations that account for the
expansion of the universe as well as for reactions that change the
total number of particles for the species of interest. If these
reactions freeze out when the temperature is not too much smaller than
the mass of the particle, there will be associated a significant
energy density that remains today that may account for the observed DM
in the universe.

From the potential in Eq.~(\ref{Lag2}), we can determine the reactions
that change the total number of particles, which may be classified as
follows:
\begin{enumerate}[\rm (\it a~\!\rm )]
\item {\it Annihilation into SM}:
\begin{eqnarray}
 &&\eta \eta  \rightarrow  t{\bar t}, hh, ZZ, WW, b\bar{b} \dots , \nonumber \\
 &&\chi \chi^*  \rightarrow   t{\bar t}, hh, ZZ, WW, b\bar{b} \dots   ~. \nonumber
\end{eqnarray}
\item {\it Semi-annihilation}:
\begin{eqnarray}
&&\chi \chi   \rightarrow h \chi^*,  \qquad
\chi h   \rightarrow \chi^* \chi^* ~. \nonumber 
\end{eqnarray}
\item {\it DM conversion}:
\begin{eqnarray}
& \eta \chi \rightarrow \chi^* \chi^*, \qquad
 \eta \chi^* \rightarrow \chi \chi, \qquad
\chi \chi   \rightarrow \eta \chi^*, \qquad & \nonumber  \\
&\eta \eta \rightarrow \chi\chi^*, \qquad  
 \chi \chi^* \rightarrow \eta \eta. & \nonumber
\end{eqnarray}
\item {\it  Late decay}:
\begin{eqnarray}
 &&\eta   \rightarrow 3 \chi,~3 \chi^*.  \nonumber 
\end{eqnarray}
\end{enumerate}
The reactions listed above constitute the relevant processes that
change the number of $\eta$ and $\chi$ particles; analogous reactions
hold for $\chi^{*}$.  (Annihilation into three body final states can
be important right below the $WW$ threshold~\cite{yaguna}.)  We see
that the $D_3$ symmetry allows for the non-canonical processes of
semi-annihilation ({\it b~\!}) and DM conversion ({\it c~\!}), studied
recently in Refs.~\cite{Hambye,semiA}, as well as a
late decay scenario ({\it d~\!})~\cite{Fairbairn}.  $CP$~invariance of
the potential (\ref{potential}) implies $n_{\chi^*} = n_\chi$ so that
we only need to solve the Boltzmann equation for $n_\chi$; the total
relic density is then $n_{\chi + \chi^*} = 2 n_\chi$.  Furthermore,
processes (\it b~\!\rm ), ({\it c~\!}), and ({\it d~\!})  couple the
Boltzmann equations for $\eta$ and $\chi$.

We have performed a general analysis of the Boltzmann equations
following the usual treatments of
Refs.~\cite{Gondolo1,Gondolo2,DarkSUSY,Luty}, with modifications for
semi-annihilation and DM conversion; a summary is presented in
Appendix~A.  In the results presented below, we obtain the relic
densities of $\eta$ and $\chi$ by numerically integrating the coupled
set of equations.  We now present a detailed survey of the cosmology,
emphasizing the novel aspects that are a result of the $D_3$ symmetry.

An immediate consequence of the non-Abelian symmetry is that since
there are two stable DM candidates for $m_\eta < 3 m_\chi$, the
individual relic abundances are not fixed by the WMAP measurement of
the DM relic density, but rather the sum is fixed: $\Omega_\eta h^2 +
\Omega_{\chi+\chi^*} h^2 = \Omega^{\rm WMAP}_{DM} h^2= 0.1126 \pm
0.0036$ \cite{wmap}. This means that different choices for a certain
combination of parameters will fit the WMAP measurement while shifting
the fractional abundances of $\eta$ and $\chi$.

\paragraph{Annihilation into SM.}

The first class of reactions ({\it a~\!}) in which $\eta$ and $\chi$
annihilate into light SM particles is common to any scalar DM model
with a Higgs portal interaction.  In the case $m_{\chi,\eta} \ll m_h$,
the thermally averaged annihilation cross section may be written as
\cite{singlet1,singlet2,singlet3}
\begin{equation}
 \langle \sigma v \rangle_{ii \rightarrow X_{SM}} \simeq 
\frac{4 \alpha_i^2 v^2}{(4m_i^2 - m_h^2)^2 + m_h^2 \Gamma_h^2}
\frac{\widetilde\Gamma_{i}}{m_i},
\end{equation}
where $i=1~\!\!(2)$ corresponds to $\eta$~\!\!($\chi$) and $X_{SM}$
denotes all kinematically allowed final states ({\em e.g.}, $X_{SM}=
ZZ, WW, b\bar{b} \dots$ for $m_i > m_Z$); $v$ in the nominator of the
right hand side stands of course for the Higgs expectation value
in~(\ref{vac}).  Furthermore, $\Gamma_h$ denotes the decay width of
the SM Higgs boson (including decays to the new dark scalars if
kinematically accessible) and we have defined ${\widetilde\Gamma_{i }}
\equiv \Gamma_{h^* \rightarrow X_{SM}}(m_{h^*} = 2 m_i)$ to denote the
decay width of a virtual SM Higgs of mass $m_{h^*}= 2 m_i$.  For
scalars above the weak scale, $m_{\eta,\chi} \gg m_h$, the
annihilation cross section becomes
\begin{equation}
\langle \sigma v \rangle_{ii \rightarrow X_{SM}} \simeq \frac{\alpha_i^2}{4 \pi m_i^2}.
\label{sigvSMH}
\end{equation}

For the lightest of the DM species, annihilation into SM final states
is particularly important since the DM conversion processes ({\it
  c~\!}) are likely suppressed by kinematics and Boltzmann statistics.
Consider, for concreteness, the case $m_\eta < m_\chi$ for which
$\eta$ dominantly depletes via annihilation into the SM final
states. This implies a sizable Higgs portal coupling $\alpha_1$ in
order to obtain an acceptable relic density. For example, for $m_\eta
\ll m_h$ and below the $s-$channel resonance at $m_\eta = m_h/2$, the
dominant annihilation channel is $\eta \eta \rightarrow b
\overline{b}$, with cross section $\langle \sigma v \rangle_{ \eta
  \eta \rightarrow b \overline{b} } \simeq 3 \alpha_1^2 m_b^2 / \pi
m_h^4$. One then finds a minimal coupling $\alpha_1 \gtrsim 0.2 \times
(m_h/ 120~{\rm GeV} )^2 $. Also in the opposite case, when $\eta$ is
heavier than the SM Higgs $h$ (but still lighter than $\chi$),
the annihilation cross section~(\ref{sigvSMH})
dictates a minimal coupling $\alpha_1 \gtrsim 0.1 \times (m_\eta
/600~{\rm GeV})$. Only near resonance, $m_\eta = m_h/2$, can the
coupling $\alpha_1$ be much smaller while still allowing for
sufficient depletion of $\eta$ particles.  Conversely, the heavier
species $\chi$ need not couple strongly to the Higgs portal since the
DM conversion processes ({\it c~\!}) are efficient for~$m_\chi >
m_\eta$.

\paragraph{Semi-annihilation.}

\begin{figure}
\centering
\includegraphics[width=\columnwidth]{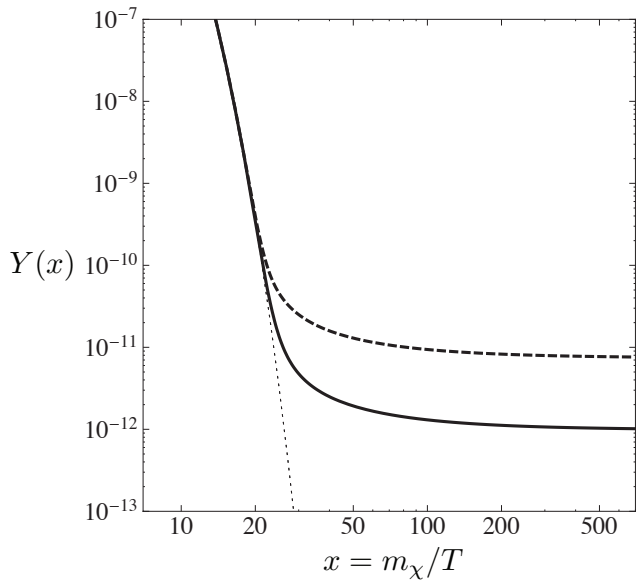}
\caption{{\it Semi-annihilation}: The comoving density $Y(x)$
  vs. $x=m_\chi/T$ for $\chi$. The thin dotted line shows the
  equilibrium value $Y^{eq}(x)$, while the thick lines show the
  evolution of the comoving density. We take $m_\chi = 200$ GeV,
  $m_h=120$ GeV, $\alpha_2 = 0.016$, and $\alpha_1 = \alpha_3
  =\alpha_4 = 0$ while assuming $m_\eta \gg m_\chi$.  With $\mu_1 = 0$
  (dashed) $\chi$ freezes out too early and overcloses the universe
  whereas with $\mu_1 = 1000$ GeV (solid) $\chi$ freezes out later so
  that the relic abundance is $\Omega_{\chi+\chi^*} = \Omega^{\rm WMAP}_{DM}$
  despite such a small value of the Higgs portal coupling
  $\alpha_2$. This illustrates that semi-annihilation allows for
  efficient depletion of $\chi$ if the cubic coupling $\mu_1$ is
  sizable when $m_\chi > m_h$.  }
\label{one}
\end{figure}

Consider now the case when $\chi$ is lighter than $\eta$.  The Higgs
portal coupling $\alpha_2$ should now be sizable to give $\chi$ the
right relic abundance.  In addition, the semi-annihilation processes
({\it c~\!}) become available for $m_\chi > m_h$.  Indeed, this may be
the dominant process for which one can obtain a semi-analytic solution
to the relic abundance of~$\chi$~\cite{semiA}:
\begin{equation}
\Omega_{\chi +\chi^*}h^2  \simeq 2 \times \frac{ 1.07 \times 10^9 x_F}{g_*^{1/2} M_P 
[ \frac{1}{2} \langle \sigma v \rangle_{\chi \chi \rightarrow \chi^* h} ] } ;
\label{approxSemi}
 \end{equation}
 $g_*$ denotes the effective number of relativistic degrees of freedom
 at the time of the freeze-out, and the Planck scale is $M_{P} = 1.22
 \times 10^{19}\,\GeV$.  The decoupling temperature $T_F$ can be
 determined upon solution of $x_F\simeq \log\{0.038 c(c+1)[
 \frac{1}{2} \langle \sigma v \rangle_{\chi \chi \rightarrow \chi^* h}
 ] m_\chi M_P g_*^{-1/2} x_F^{-1/2} \} $ with $c=\sqrt{2}-1$, where
 $x=m_\chi/T$.  In the $D_3$ model the thermally averaged
 semi-annihilation cross section is given by
\begin{equation}
  \langle \sigma v \rangle_{ \chi \chi \rightarrow h \chi^*} \simeq \frac{3 \alpha_2^2 \mu_1^2 v^2}{32 \pi m_\chi^6}.
\label{svsemi}
\end{equation}
We see from Eq.~(\ref{svsemi}) that both the Higgs portal coupling
$\alpha_2$ and the cubic coupling $\mu_1$ must be nonzero for
semi-annihilation to play a role. Because $\alpha_2 \neq 0$,
annihilation into SM final states ({\it a~\!}) can occur as well and
will in general dominate unless the coupling $\mu_1$ is relatively
large. Furthermore, because of the steep falloff of the
semi-annihilation cross section (\ref{svsemi}) with increasing
$m_\chi$, the process is most effective when $\chi$ is not much
heavier than the SM Higgs, $m_{\chi}\gtrsim m_{h}$.  The effect of
semi-annihilation is illustrated in Fig.~\ref{one}, which shows the
evolution of the comoving density $Y \equiv n/s$, where $s$ is the
entropy density.

\paragraph{DM Conversion.}

We next consider the processes of DM conversion ({\it d~\!}), which
consist of $2 \rightarrow 2$ reactions that convert DM of one species
into another. There are two types of reactions within class ({\it
  d~\!}) that may be distinguished. The reactions in the first line of
({\it d~\!}) may be regarded as ``dark'' semi-annihilation ({\em
  e.g.}, $\chi \chi \rightarrow \eta \chi^*$, etc.), in which the
number of DM particles of a particular species changes by one
unit. Such processes are governed by the nontrivial interaction
predicted by the $D_3$ symmetry in Eq.~({\ref{potential}}) with
coupling $\alpha_4$. The reactions in the second line of ({\it d~\!})
are regarded as ``dark'' annihilation ({\em e.g.}, $\chi \chi^*
\rightarrow \eta \eta$ and the inverse process), which change the
number of each dark species by two units. The dark annihilation
processes can be mediated by the Higgs portal when $\alpha_{1,2}$ are
sizable, or via the direct scalar coupling $\alpha_3$ in
Eq.~({\ref{potential}}).

The DM conversion processes that reduce the number of the heavier DM
species, be it $\eta$ or $\chi$, are kinematically favorable and hence
very efficient at depleting such heavy particles. Thus DM conversion
may have a dramatic influence on the eventual number density of the
heavier species.  Unless the masses of DM particles are nearly
degenerate, the relic abundance of the lighter DM species will not be
influenced.

\begin{figure}
\centering
\includegraphics[width=\columnwidth]{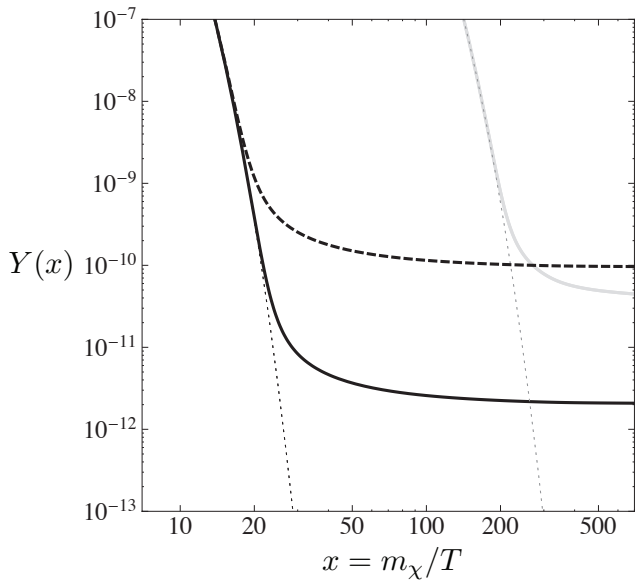}
\caption{{\it DM Conversion}: The comoving density $Y(x)$
  vs. $x=m_\chi/T$ for $\eta$ (gray) and $\chi$ (black). The thin
  dotted lines show the equilibrium value $Y^{eq}(x)$, while the thick
  lines show the evolution of the comoving density. We take $m_\eta=5$
  GeV, $m_\chi = 50$ GeV, $m_h=120$ GeV, $\alpha_1 = 0.45$, and $\mu_1
  = \alpha_2 = \alpha_3 = 0$.  With $\alpha_4 = 0.01$ (dashed black)
  $\chi$ freezes out too early and overcloses the universe whereas
  with $\alpha_4 = 0.075$ (solid black) $\chi$ freezes out later so
  that the total relic abundance is $\Omega_\eta +
  \Omega_{\chi+\chi^*} = \Omega^{\rm WMAP}_{DM}$. This demonstrates
  that DM conversion is very efficient at depleting the heavier DM
  species $\chi$. Conversely, the relic abundance of the lighter
  species $\eta$ is the same for both values of $\alpha_4$ and
  therefore not affected by DM conversion.  }
\label{two}
\end{figure}

Let us illustrate the effect of DM conversion by assuming for the
moment that $m_\chi \gg m_\eta$ for which there are two stable DM
species. As discussed above, $\eta$---being the lighter state---must
have a sizable Higgs portal coupling $\alpha_1$.  On the other hand,
$\chi$ may deplete via DM conversion processes. For example, if the
dominant process is $\chi \chi \rightarrow \eta \chi^*$, the relic
abundance may be estimated from~(\ref{approxSemi}) with the
replacement $ \langle \sigma v \rangle_{\chi \chi \rightarrow \chi^*
  h}\to \langle \sigma v \rangle_{\chi \chi \rightarrow \eta \chi^*
}$. In our minimal $D_3$ model the ``dark'' semi annihilation cross
section is found to be
\begin{equation}
  \langle \sigma v \rangle_{\chi \chi \rightarrow \eta \chi^*} \simeq \frac{3 \alpha_4^2}{128 \pi m_\chi^2}.
\label{svDMconv}
\end{equation}
One needs in this case a minimal coupling $\alpha_4 \gtrsim 0.05
\times (m_\chi/50\,{\rm GeV})$. In contrast to semi-annihilation into
the SM Higgs (\ref{svsemi}), ``dark'' semi-annihilation
(\ref{svDMconv}) is quite efficient and requires only a modest
coupling $\alpha_4$ to significantly influence the relic abundance of
$\chi$.  We illustrate the effect of DM conversion in
Fig.~(\ref{two}), focusing on ``dark'' semi-annihilation. A similar
discussion applies to ``dark'' annihilation ({\em e.g.}, $\chi\chi^*
\rightarrow \eta\eta$, etc).

\paragraph{Late decay of $\eta$.}

The final novel aspect of the cosmology in the minimal $D_3$ model is
the possibility of the late decay of $\eta$. If $m_\eta > 3 m_\chi$,
the state $\eta$ will decay via $\eta \rightarrow 3 \chi, 3\chi^*$ due
to the coupling $\alpha_4$.  In the limit $m_\eta \gg 3 m_\chi$, the
total width of $\eta$ is approximately given by
\begin{equation}
 \Gamma_\eta \simeq \frac{\alpha_4^2 m_\eta}{1536 \pi^3} ,
\end{equation}
and which implies a lifetime of
\begin{equation}
\label{eq:etalifetime}
\tau_\eta  \simeq  10^{-8}\, {\rm s}  \times \left( \frac{10^{-7}}{\alpha_4} \right)^{2} \left(\frac{100 \,{\GeV}}{m_\eta}\right).
\end{equation}

\begin{figure}
\centering
\includegraphics[width=\columnwidth]{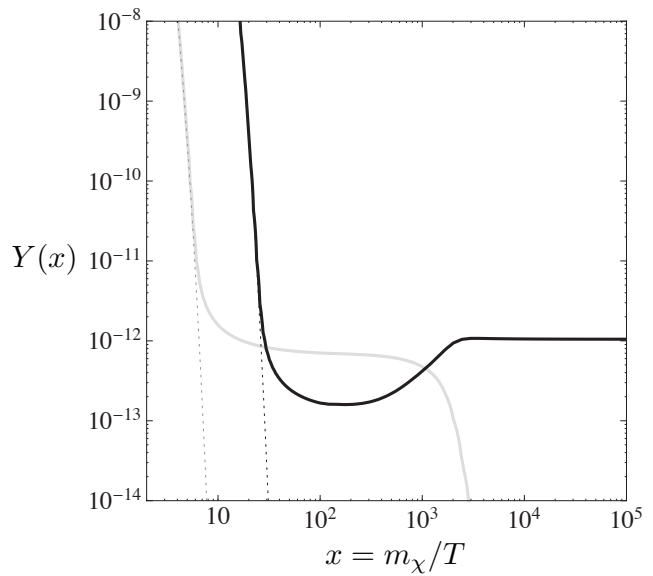}
\caption{ {\it Late decay}: The comoving density $Y(x)$
  vs. $x=m_\chi/T$ for $\eta$ (gray) and $\chi$ (black). The thin
  dotted lines show the equilibrium value $Y^{eq}(x)$, while the thick
  lines show the evolution of the comoving density. We take
  $m_\eta=800$ GeV, $m_\chi = 200$ GeV, $m_h=120$ GeV, $\alpha_1 =
  0.11$, $\alpha_2 = 0.15$, $\alpha_4 = 10^{-9}$ and $\mu_1 = \alpha_3
  = 0$, leading to a long lifetime $\tau_\eta \sim 40 \mu$s. Initially
  $\chi$ is depleted through efficient annihilation into SM particles,
  whereas $\eta$ freezes out with a significant density. Later,
  $\eta$ decays and repopulates $\chi$, $\chi^*$ which then comprise
  the observed DM.  }
\label{three}
\end{figure}

For small values of $\alpha_4$ 
(which are technically natural), $\eta$ becomes long-lived in the
sense that it decays after chemical decoupling, $\tau_\eta \gg
t_{\eta,F}$. The cosmic time at freeze-out reads $t_{\eta,F} \simeq
10^{-8}\,{\rm s} \times(m_\eta /100\,{\rm GeV})^{-2}
(x_{\eta,F}/20)^2$, where $x_{\eta}\equiv m_\eta/T$ and $T_{\eta,F}$
is the decoupling temperature. For a given $m_{\eta}$, it is then easy
to read off the maximum values of $\alpha_4$ from
(\ref{eq:etalifetime}) so that the late-decay condition $\tau_\eta \gg
t_{\eta,F}$ holds.  Clearly, the decays $\eta\rightarrow 3\chi,
3\chi^*$ can then significantly contribute to the relic abundance
$\Omega_{\chi+\chi^*}h^2$---provided that $\chi$-annihilation has
frozen out.  In the limit where $\chi$ is initially depleted, late
$\eta$-decays lead to a non-thermal $\chi$-population with relic DM
abundance
\begin{equation}
\Omega_{\chi+\chi^*} h^2 = 3 \frac{m_\chi}{m_\eta} \Omega_\eta h^2,
\end{equation}
where $\Omega_\eta h^2$ represents the would-be relic abundance of
$\eta$ particles today, had they not decayed. The late decay scenario
is illustrated in Fig.~\ref{three}.

For $\tau_{\eta} > 1\,\mathrm{s}$, the late-decay scenario
\textit{(d)} is cosmologically constrained. The main decay mode
$\eta\to 3\chi^{(*)}$ is inevitably accompanied by $\eta\to
3\chi^{(*)}+ h^{*}$ where the virtual $h^{*}$ decays into
kinematically accessible final states. This leads to electromagnetic
and hadronic energy injection into the primordial plasma. If the decay
happens during or after Big Bang nucleosynthesis (BBN) the induced
spallation reactions of the light elements can spoil the concordance
of BBN abundance predictions with their observationally inferred
values (for a review see, \textit{e.g.}, \cite{Pospelov:2010hj}.)
Since energy depositions as small as $\mathcal{O}(\mathrm{few}\,\MeV)$
per baryon can probed in BBN~\cite{BBNconstr}, branching fractions of
the $3\chi+ h^{*}$ decay mode of, say, $\mathcal{O}(10^{-4})$ are
still constrained. To circumvent this issue as well as potential warm
DM constraints due to the free streaming of the decay products $\chi$,
one can simply impose $\tau_{\eta}<1\,\mathrm{s} $. For a
kinematically unsuppressed decay, this requires $ |\alpha_4| \gtrsim
10^{-11}\sqrt{100\,\GeV/m_{\eta}}$.

Finally, we note that in the limit $\alpha_4 \rightarrow 0$, $\eta$
becomes stable even for $m_\eta > 3 m_\chi$. This is easily understood
since in this limit the Lagrangian displays a $Z_2$ symmetry $\eta
\rightarrow - \eta$ which renders $\eta$ stable.

\paragraph{\bf Direct detection phenomenology.}

The minimal $D_3$ model of DM can be efficiently probed by direct
detection experiments. Furthermore, compared to the canonical $Z_2$
scalar singlet model, the non-Abelian $D_3$ model offers several
distinct direct detection signatures which can be observed by current
and next generation experiments. We now examine the constraints and
future prospects offered by these experiments.

We use the standard formalism \cite{Lewin} to predict the nuclear
recoil rates from elastic scattering of $\eta$ and $\chi$ .  The
differential event rate is given by
\begin{equation}
 \frac{dR_i}{dE_R} =N_T \frac{\rho_i}{m_i}
 \int_{|\bm{v}|\geq v_{\mathrm{min}}} d^3\bm{v} \, v f(\bm{v},\bm{v}_\mathrm{e}) \frac{d\sigma_i}{d E_R},
\label{diffrate}
\end{equation}
where $i=1~\!\!(2)$ corresponds to $\eta$~\!\!($\chi$), $N_T$ is the
number of target nuclei per unit detector mass, $\rho_i$ is the local
DM mass density for a given species, and $f(\bm{v},\bm{v}_\mathrm{e})$
is the DM velocity distribution taken to be Maxwellian and truncated
at an escape velocity of $v_{\mathrm{esc}}= | \bm{v} +
\bm{v}_\mathrm{e} | = 600\,$km/s; $v_{\mathrm{min}}$ is the minimum
velocity required to cause a nuclear recoil with energy $E_R$ and
$\bm{v_{\mathrm{e}}}$ and $\bm{v}$ are the velocity of the earth in
the galactic frame and the velocity of the DM particle in the earth's frame,
respectively.
The theoretical rate in Eq.~(\ref{diffrate}) is corrected to account
for potential quenching of the recoil signal as well as for finite
detector resolution, efficiency, and acceptance to obtain a prediction
for the observed rate.  The total number of events in a certain energy
interval is then found by integrating the rate and multiplying by the
raw exposure.

The microscopic physics of our $D_3$ model enters into the scattering
cross section in Eq.~(\ref{diffrate}). The scalar representation of
$D_3$ gives 
rise to spin-independent DM-nucleus scatterings. The
differential cross section is conventionally expressed as
\cite{review},
\begin{equation}
  \frac{d \sigma_i}{d E_R} = \frac{m_N}{2 v^2} \frac{\sigma^{(i)}_{n} }{\mu_n^2}
  \left[\frac{f_p Z +f_n(A-Z)}{f_{n}}\right]^2 F^2(E_R),
\label{sigmanucleon}
\end{equation}
where $m_N$ is the mass of the target nucleus, $\mu_n$ is the
DM-nucleon reduced mass, and $F^2(E_R)$ is the nuclear form factor; we
take the Helm form factor following~\cite{Lewin} with nuclear skin
thickness $s=0.9$\,fm. Spin-independent elastic scattering of the DM
particles $\eta$ and $\chi$ with nuclei is mediated via a $t-$channel
exchange of the SM Higgs boson and therefore governed by the couplings
$\alpha_1$ and $\alpha_2$, respectively.  The DM-nucleon cross section
is given by
\begin{equation}
\sigma^{(i)}_n = \frac{\alpha_i^2 f_n^2 \mu_n^2}{\pi m_i^2 m_h^4},
\label{sign}
\end{equation}
where 
$f_n = \sum_q \langle n | m_q \bar n n |n \rangle \simeq 0.52$ GeV is
the effective DM-nucleon coupling~\cite{Ellis}.

Because there are two species of DM in our model, the local density of
each species is typically not equal to $\rho_0 \approx 0.3$ GeV
cm$^{-3}$.  As in Ref.~\cite{dualDD} we assume that the local density
of each DM particle is proportional to its cosmological abundance,
\textit{i.e.},
\begin{equation}
\rho_i = \frac{\Omega_i}{\Omega^{\rm WMAP}_{DM}} \rho_0.
\label{rhoi}
\end{equation}
For a given set of model parameters, we compute the relic density
$\Omega_i$ of each species by integrating the Boltzmann equations,
find the local DM mass density $\rho_i$ according to Eq.~(\ref{rhoi}),
and calculate the nuclear scattering rate (\ref{diffrate}).

\paragraph{Experiments.}

In the following discussion we shall focus on the liquid xenon
experiments in anticipation of the upcoming XENON100 one-year data
release as well as the prospect of future ton-scale experiments which
have the potential to probe large regions of the $D_3$ model parameter
space. Furthermore, past data sets from XENON10~\cite{Angle:2007uj,Angle:2009xb} and XENON100~\cite{Aprile:2010um} yield
current limits on spin-independent scattering that are competitive
with other experiments, such as CDMS
II~\cite{Ahmed:2009zw,Ahmed:2010wy}.
The XENON experiments use a prompt scintillation light signal (S1) and
a delayed ionization signal (S2) to discriminate between nuclear and
electron recoil events. Recently, there has been a lot of debate on
the scintillation yield for nuclear recoils,
$\mathcal{L}_{\mathrm{eff}}$, as well as on the stochasticity of the
S1 signal for lowest $E_{R}$. Since $\mathcal{L}_{\mathrm{eff}}$
defines the detector threshold, this quantity is of importance when
inferring $\sigma_n$-limits for small DM masses
$\mathcal{O}(10\,\GeV)$~\cite{Leff,Sorensen:2010hq}. We use the results
of the detailed study~\cite{Sorensen:2010hq} to account for
resolution, efficiency, and acceptance of the
XENON10~\cite{Angle:2007uj} and XENON100~\cite{Aprile:2010um}
detectors where we employ the most conservative assumption on
$\mathcal{L}_{\mathrm{eff}}$ (yielding the weakest limits on
$\sigma_n$.)  Constraints are obtained by Yellin's maximum gap
method~\cite{Yellin:2002xd} which accounts for the $\mathcal{O}(10)$
events observed in the XENON10 re-analysis~\cite{Angle:2009xb} of its
316\,kg$\times$days data sample. The XENON100 collaboration has
published an analysis of an 11\,day run yielding a raw exposure of
447\,kg$\times$days~\cite{Aprile:2010um}; no candidate events were
observed.

In addition, a data release from a one year run of XENON100 is
expected shortly. To estimate the reach of this data set we assume an
exposure of 100 live days, collected with their fiducial detector mass
of 30\,kg.  Lastly, for the sake of exploring future sensitivity of a
ton-scale liquid xenon experiment, following Ref.~\cite{Pato:2010zk}
we assume less than one background event; we take a (moderate) raw
exposure of 1\,ton$\times$yr.  We mimic energy resolution and
efficiency of the detector by employing the same
analysis~\cite{Sorensen:2010hq} as used for the XENON10 detector
above; this implies a more optimistic assumption on the low recoil
acceptance then had we used the similar analysis for XENON100.

As we now show, large portions of the parameter space in the $D_3$
model have been or will be probed by liquid xenon experiments. We will
describe the constraints and future sensitivities in detail for two
novel scenarios that may occur in the $D_3$ model: 1) the late-decay
scenario ($m_\eta > 3 m_\chi$), and 2) the two-component DM scenario
($m_\eta < 3 m_\chi$).

\paragraph{Constraints for $m_\eta > 3 m_\chi$.}

If $\eta$ is more than three times as heavy as $\chi$, it is unstable,
dictating that $\chi$ accounts for all of the cosmological DM. Without
making assumptions about the lifetime of $\eta$ (or equivalently the
coupling $\alpha_4$), the Higgs portal coupling $\alpha_2$ can take a
range of values and still be consistent with the relic abundance
constraint. The minimum allowed value of $\alpha_2$ is determined by
the requirement that $\chi$ does not overclose the universe upon
freezeout. However $\alpha_2$ can in fact be much larger than this
minimum value because late decays of $\eta$ can replenish an initially
depleted population of $\chi$ particles.

\begin{figure}
  \centering
\includegraphics[width=\columnwidth]{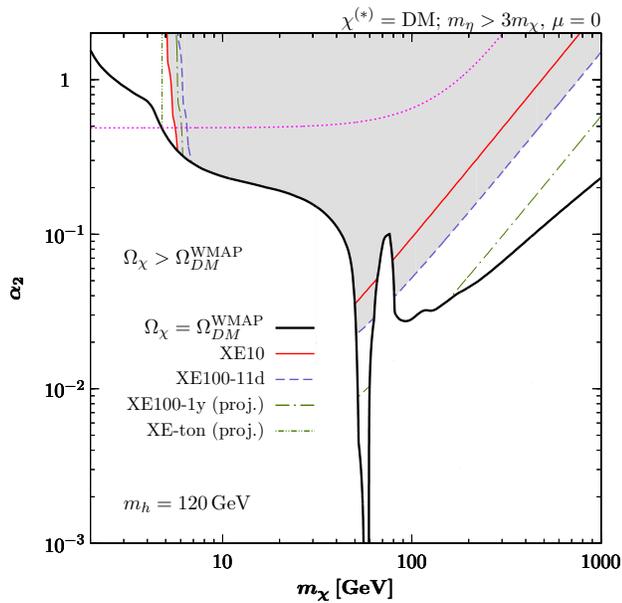}%
\caption{ {\it Direct detection constraints}: Shown is the
  $m_\chi$--$\alpha_2$ plane for $m_{\eta}>3 m_{\chi}$ and
  $\mu=0$. Along the thick solid (black) line $\Omega_{\chi+\chi^*}=
  \Omega^{\rm WMAP}_{DM}$ holds. The region below is excluded,
  $\Omega_{\chi+\chi^*} > \Omega^{\rm WMAP}_{DM}$. Only $\chi^{(*)}$
  is stable so that above the solid black line $\eta\to 3\chi^{(*)}$
  populates $\chi^{(*)}$ such that $\rho_{\chi}=\rho_0$. The gray
  shaded region is excluded by current direct detection
  experiments. As labeled, the lines depict the individual experiments
  and show their current (projected) sensitivity for exclusion. For
  $m_{\chi} \sim m_h/2 = 60\,\GeV$ primordial annihilation can proceed
  resonantly via Higgs exchange. The horizontal dotted line shows a
  vacuum stability constraint for $\lambda_3 = 2 $; values above are
  excluded.}
\label{amplane1}
\end{figure}

In Fig.~\ref{amplane1} we show the 
constraints in the $m_\chi$--$\alpha_2$ plane for the case $m_{\eta}>3
m_{\chi}$ and $\mu = 0$ (so that semi-annihilation $\chi\chi
\rightarrow \chi^* h$ is absent).  Along the thick solid line the
relic abundance of~$\chi$ is in concordance with that inferred from
observation, $\Omega_{\chi+\chi^*}= \Omega^{\rm WMAP}_{DM}$. The
region below is excluded since $\Omega_{\chi+\chi^{*}} > \Omega^{\rm
  WMAP}_{DM}$. In the region above, the decay $\eta\to 3\chi^{(*)}$
must source the relic abundance to obtain $\Omega_{\chi+\chi^*}=
\Omega^{\rm WMAP}_{DM}$ and thus a viable cosmology. We see that,
except near the resonance, $2 m_\chi \simeq m_h = 120\,\GeV$, the
Higgs portal coupling $\alpha_2$ must be sizable.  The shaded region
is excluded due to measurements of past direct detection
experiments. In particular, the light DM region with $m_\chi \lesssim
10~\GeV$ is most constrained by XENON10 (as well as by the CDMS-II low
threshold analysis~\cite{Ahmed:2010wy}), while heavier DM is most
constrained by the XENON100 first run with 11~live days (as well as
the final exposure of CDMS II~\cite{Ahmed:2009zw}).  We also show the
projections for the upcoming release of XENON100 one-year data, as
well as a future ton-scale xenon experiment. While the former is not
expected to improve upon the XENON10 low-mass limit, the latter will
not only improve upon this limit but will essentially probe all values
of $m_{\chi}$ that lie to the right of the Higgs resonance region.
Finally, the dotted horizontal line shows the requirement that the
electroweak vacuum~(\ref{vac}) is a global minimum; values of
$\alpha_2$ above that line will make the vacuum configuration
metastable. The bound depends on the quartic coupling $\lambda_3$
(here $\lambda_3 = 2$) and also on the sign of~$\alpha_2$.

\begin{figure}
  \centering
\includegraphics[width=\columnwidth]{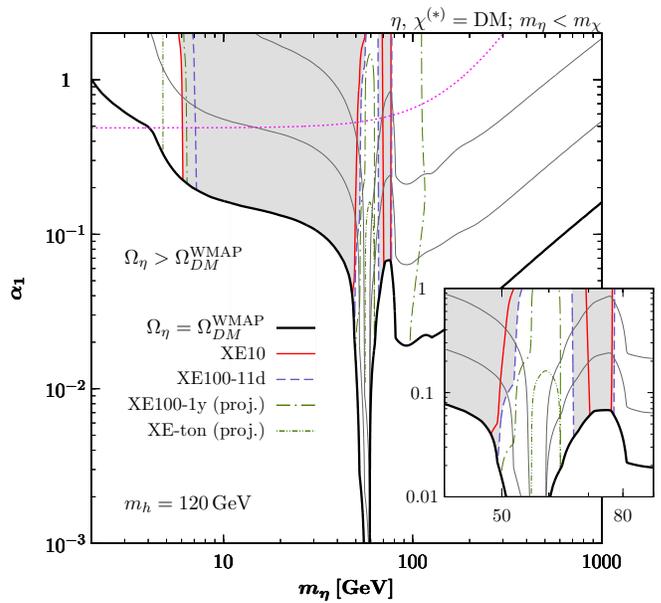}%
\caption{ {\it Direct detection constraints}: Shown is the
  $m_\eta$--$\alpha_1$ plane for $m_{\eta}<m_{\chi}$. Along (below) the
  thick solid line $\Omega_{\eta}= \Omega^{\rm WMAP}_{DM}$
  ($\Omega_{\eta} > \Omega^{\rm WMAP}_{DM}$) holds. The thin gray
  lines parallel and above are contours with a fraction of 0.1 and
  0.01 of $\Omega^{\rm WMAP}_{DM}$ (bottom to top.)  The near-to
  vertical lines depict the individual experiments and show their
  current (projected) sensitivity for exclusion. The Higgs resonance
  region is shown more closely in the inlay.}
\label{amplane2}
\end{figure} 

\paragraph{Constraints for $m_\eta < 3 m_\chi$.}

If $m_\eta < 3 m_\chi$, there are two DM species since the decay $\eta
\rightarrow 3\chi^{(*)}$ is kinematically forbidden. The individual
relic abundances of $\eta$ and $\chi^{(*)}$ are not constrained by
WMAP, and so it may appear that further model assumptions are
necessary to constrain the two-component scenario. In fact, robust
direct detection constraints exist for the lightest DM species
independent of the fractional composition of the cosmological dark
matter.  For the sake of discussion, let us assume that $\eta$ is the
lightest DM particle, $m_\eta < m_\chi$. As discussed in detail above,
the only kinematically favorable annihilation channel for $\eta$ is
into the light SM states~(\ref{sigvSMH}). This means that $\eta$ must
have a minimum Higgs portal coupling $\alpha_1$. Because the same
coupling $\alpha_1$ controls the DM-nucleon scattering cross section
(\ref{sign}), direct detection experiments constrain large portions of
the $m_\eta$--$\alpha_1$ parameter space.

In Fig.~\ref{amplane2} we show constraints in the $m_\eta$--$\alpha_1$
plane for the case $m_{\eta}< m_{\chi}$.  The lines as labeled are
analogous to the ones shown in the previous Fig.~\ref{amplane1}. This
time, however, the regions between $6\,\GeV \lesssim m_\eta \lesssim
50\,\GeV$ and $70\,\GeV \lesssim m_\eta \lesssim 80\,\GeV $ are
excluded for {\it all} values of $\alpha_1$. This may seem
counterintuitive at first sight, since as $\alpha_1$ increases the
relic abundance decreases: $\Omega_\eta \sim 1/\langle \sigma v
\rangle_{\eta\eta \rightarrow X_{SM} } \propto \alpha_1^{-2}$ so that
$\rho_\eta \propto \alpha_1^{-2}$ by Eq.~(\ref{rhoi}). This can also
be seen by the contour lines parallel to the thick solid line which
now show decreasing fractions 0.1 and 0.01 (from bottom to top) of the
DM abundance. For increasing $\alpha_1$ there is, however, a
corresponding increase in the DM-nucleon scattering cross
section~(\ref{sign}): $\sigma^{\eta}_n \propto \alpha_1^2$. Since
$dR_\eta/dE_R \propto \rho_\eta \sigma^{\eta}_{n}$ the dependence on
$\alpha_1$ cancels out in the nuclear scattering
rate~(\ref{diffrate}). This explains why the various direct detection
constraints as labeled are now near-to vertical lines.

While the constraints on the lightest DM particle (assumed to be
$\eta$ for the discussion) are not strongly dependent on the
assumptions made about other model parameters, the same is not true
for the heavier species (here assumed to be $\chi$). This is because
the coupling $\alpha_2$ governing the interactions of $\chi$ with the
SM is largely unconstrained by the cosmology, since $\chi$ may
efficiently deplete through DM conversion processes ( {\em e.g.},
$\chi \chi \rightarrow \chi^* \eta$, etc.). Hence further assumptions
must be made about other couplings in the Lagrangian (\ref{potential})
to make concrete statements about the heavier DM component.

\paragraph{Discovering  two-component dark matter.}

The two-component nature of DM for $m_\eta < 3 m_\chi$ is one of the
most intriguing features of the non-Abelian $D_3$ discrete
symmetry. Naturally, the question arises: can one discriminate between
a single-component and a two-component DM scenario at future direct
detection experiments? Here we will attempt to provide a quantitative
answer to this question.  Our analysis will be based on the shape of
the nuclear recoil spectrum, which may exhibit distinct features
depending on whether one or two components are scattering in the
detector.

\begin{figure}
\centering
\includegraphics[width=\columnwidth]{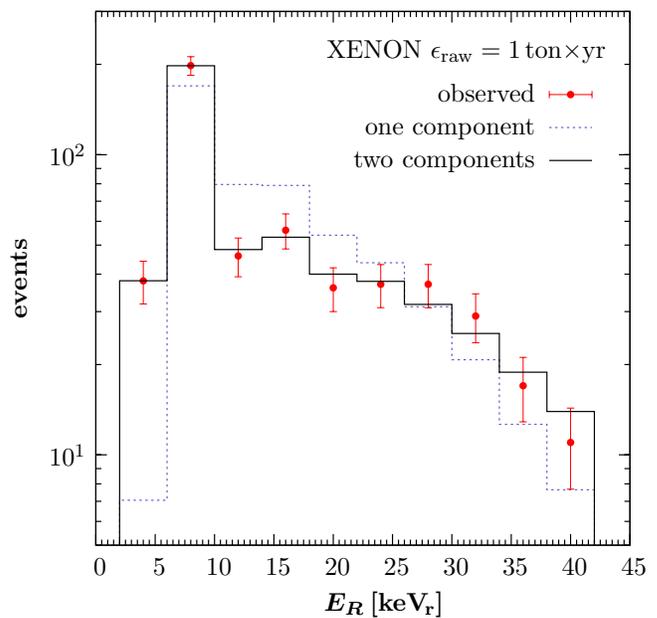}
\caption{ {\it Recoil spectrum:} The dots show a Monte Carlo generated
  recoil spectrum of the two-component DM scenario with input values
  $m_{\eta}=5\,\GeV$, $\alpha_1 = 0.45$, $m_{\chi} = 200\,\GeV$,
  $\alpha_2= 0.065$, and $\rho_{1,2} = \rho_0/2$ for a ton-scale
  liquid Xe detector with raw exposure of
  $1\,\mathrm{ton}\times\mathrm{yr}$. The dashed line (blue) shows the
  attempt to explain the data with a generic single DM particle. The
  solid line (black) is obtained from a DM model with two-components;
  see main text for details.}
\label{four}
\end{figure}

To answer the question posed above we Monte Carlo generate a sample
recoil spectrum for a ton-scale liquid xenon experiment with
$m_{\eta}=5\,\GeV$ and $m_{\chi}=200\,\GeV$. We further set $\alpha_1
= 0.45$, $\alpha_2 = 0.065$ and $\alpha_4=0.3$ for which the dark
matter is distributed equally, $\Omega_{\eta} = \Omega_{\chi} =
\Omega^{\rm WMAP}_{DM}/2$ (note that DM conversion is used to deplete
$\chi$). This parameter point is consistent with current direct
detection constraints (see Fig.~\ref{amplane2}).  Since the XENON
experiments have a poor energy resolution, the bin-width is chosen to
be $4\,\keV_r$ with a total number of ten bins, starting from
$2\,\keV_r$.  The total number of events to be generated is then drawn
from a Poisson distribution with its mean given by the number of
theoretically expected events.  A sample ``experimental'' spectrum is
shown by the dots in Fig.~\ref{four}. The events in the first bin are
dominated by scatterings of the light state $\eta$ on the Xe
target. In the second bin both states $\eta$ and $\chi^{(*)}$
contribute equally. At higher recoil energies all events are
essentially due to~$\chi^{(*)}$.  This behavior can of course be
traced back to the favorable relation $m_{\eta}\ll m_{\chi}$ which
yields rather distinct exponential forms of $dR/dE_R$.

We now attempt to discriminate the two-component origin of the
generated spectrum from a single-component DM hypothesis. For this we minimize
the Poisson log-likelihood function
\begin{align}
\label{eq:likelihood}
  \chi_{\lambda}^2 = 2 \sum_{\mathrm{bins}\ i} \left[ N_i^{\mathrm{th}} - N_i^{\mathrm{obs}} +
    N_i^{\mathrm{obs}} \ln(N_i^{\mathrm{obs}}/N_i^{\mathrm{th}}) \right]
\end{align}
to obtain the best fit and also to assess the
goodness-of-the-fit~\cite{baker}. Here, $N_i^{\mathrm{obs}} $
($N_i^{\mathrm{th}} $) is the number of observed (theo\-retically
predicted) events in the $i$-th bin. We use the program
MINUIT~\cite{James:1975dr} to minimize (\ref{eq:likelihood}).  Note
that with the procedure outlined above, we only account for
statistical uncertainties in the event rate; a more detailed analysis
incorporating systematic uncertainties goes well beyond the present
scope.

We first attempt to fit the generated recoil spectrum with a single DM
particle.  The DM mass and spin independent DM-nucleon cross section
are used as the fitting parameters.  Using the procedure described
above, one obtains the dotted (blue) histogram in Fig.~\ref{four}.
As can already be seen by eye, the one-component fit is poor. Indeed,
from the goodness-of-fit test the single-DM hypothesis is rejected
with confidence well above~99\%.

We now turn to a two-component DM model.  In this case the recoil
spectrum depends on the DM masses $m_i$, cross sections,
$\sigma_n^{(i)}$, and local densities $\rho_i$, with $i=1,2$ and
subject to the constraint $\rho_1+\rho_2 = \rho_0$. Thus, there are in
principle five parameters to be inferred from the data of
Fig.~\ref{four}. There exist, however, degeneracies between different
parameters in the two-component spectrum so that only certain
combinations are accessible.  To understand this it is helpful to
examine the parametric dependence of the recoil
spectrum~(\ref{diffrate}) on $m_i$, $\rho_i$, and $\sigma_n^{(i)}$.
Using~(\ref{sign}), and in the limit of $|\bm{v}_{\mathrm{e}}|=0$ and
$v_{\mathrm{esc}}\to\infty$ one finds
\begin{align}
  \label{eq:param-dep-recspec}
  \frac{dR_i}{dE_R} \propto \frac{\rho_i\sigma_n^{(i)}}{m_i} \exp
  \left[ - \left( 1+\frac{m_N}{m_i} \right)^2 \frac{E_R}{2 m_N v_0^2}
  \right] ,
\end{align}
where $v_0$ is the DM velocity dispersion. Observe that 1) only the
product $\rho_i\sigma_n^{(i)} $ enters in $dR/dE_{R}$ and 2) the
exponential shape of the spectrum becomes independent of the DM mass
for $m_i \gg m_N$. From the first observation it is clear that one
cannot independently determine both $\rho_i$ and $\sigma_n^{(i)}$, but
only the product. The second observation suggests that a second
degeneracy between $\rho_i\sigma_n^{(i)} $ and $m_i$ develops once
$m_i \gtrsim m_N$. Therefore, further assumptions about the model
parameters are required if one desires to break these degeneracies.
Fixing, for example, mass and density of the heavier DM to its input
values, one then easily recovers recovers all remaining model
parameters within a ``$1\sigma$''-range or better. The result of this
fit is shown by the solid (black) histogram in Fig.~\ref{four}. 

Note that the successful determination of the parameters of the
lighter state required that at least two bins were populated. If one
tried to fit the light DM particle to a single bin, many combinations
$(m_1,\sigma_1)$ would reproduce the signal thus leading to a high
degree of degeneracy in $m_1$ and $\sigma_1$. In this regard, it is
important to note that---complementary to liquid noble gas
experiments---potential future ton-scale cryogenic detectors such as
superCDMS~\cite{Bruch:2010eq} or EURECA~\cite{eureca} may be very
powerful in disentangling the parameters of a multi-component DM
scenario. The reason lies in a much better energy resolution. For
example, Ref.~\cite{Pato:2010zk} estimates for a ton-scale Ge detector
$\sigma_{\mathrm{Ge}} < 0.5\,\keV$ ($E_R\leq40\,\keV$).  This will
allow one to obtain a more detailed picture on the spectral shape of
the signal. Moreover, a different target mass with respect to Xe may
prove most valuable when attempting to draw differential conclusions
by combining results from both detector designs.

\paragraph{\bf Discussion.}

In this paper we have investigated the simplest model in which DM is
stabilized by a non-Abelian discrete symmetry. The model is based on
the symmetry group $D_3$, which is the smallest non-Abelian finite
group. The non-Abelian nature of this theory manifests itself in the
matter content and interactions of the dark sector, which leads to
multi-component DM and a novel early universe cosmology. Robust
constraints from direct detection experiments exist for the lightest
DM species, and the two-component nature of DM can potentially be
tested at future ton-scale experiments.

While we have focused on the phenomenology relevant for direct
detection experiments, the $D_3$ model also has implications for high
energy colliders and indirect detection probes.  At colliders, the
most important consequence of the $D_3$ model, like other models in
which DM couples to the SM through the Higgs portal, is to cause the
Higgs boson to decay invisibly if kinematically allowed.  An invisibly
decaying Higgs boson can be discovered using the vector boson fusion
and $ZH$ production channels at the LHC ($\sqrt{s} = 14$ TeV) with 30
fb$^{-1}$ of data~\cite{atlas}.  Another generic DM collider probe is
a monojet produced in association with pairs of dark matter
particles~\cite{mono1,mono2}. 
In scalar singlet DM models, 
this occurs via gluon fusion
accompanied by emission of a jet.
While the Tevatron is not
sensitive to the monojet signature in this class of DM models, the LHC
could potentially probe regions with large Higgs portal couplings
$\alpha_{1,2} \sim O(1)$~\cite{mono2}.  We also note that a more
complete study of the monojet signature in Higgs portal DM models
going beyond the effective field theory approach is warranted given
that a relatively light Higgs and a top quark loop enter into the
amplitude.

A potential indirect probe of the $D_3$ model is to observe gamma rays
from annihilating $\eta$ or $\chi$ particles. In particular, Fermi-LAT
data on the isotropic gamma-ray diffuse emission~\cite{diffuse1},
depending on various astrophysical assumptions, can constrain low mass
$< 10$ GeV DM particles annihilating through the Higgs
portal~\cite{diffuse2}. For such constraints to be relevant, the
lightest component should carry most of the cosmic DM abundance, since
the gamma-ray flux scales as the square of the DM density. In
particular, light DM particles with very large Higgs portal couplings
will be depleted efficiently and therefore not subject to such
constraints. Another possibility is to observe monochromatic gamma-ray
lines via $\eta \eta, \chi \chi^*\rightarrow h^* \rightarrow \gamma
\gamma$. This has been studied in Ref.~\cite{monogamma}, and Fermi-LAT
data allows one to probe the model deep within the resonance region
$m_{\eta,\chi} \sim m_h/2$.

There are many directions for future investigations with DM and
non-Abelian discrete symmetries. Within the minimal $D_3$ model it
would be interesting to consider the case in which $\eta$ ($\chi$)
condense, leading to the spontaneous breaking of the $D_3$ symmetry to
the $Z_3$ ($Z_2$) subgroup. This will still lead to a viable model of
DM and likely will have distinct signatures compared to the unbroken
$D_3$ symmetry.  It would also be interesting to explore larger
non-Abelian finite groups, and ultimately non-Abelian discrete gauge
symmetries as the origin of DM stability.  Finally, given that
non-Abelian discrete symmetries may underlie the patterns observed in
the quark and lepton masses and mixings, it is interesting to
speculate that DM stability may intimately be connected to flavor
symmetries.

Stability on cosmological time scales constitutes one of the few and
robust guiding principles in formulation of a theory of dark matter.
Alternatives to the canonical $Z_2$ parity symmetry may lead to new
phenomena associated with dark matter, as has been clearly
demonstrated by our study of the minimal model of $D_3$ dark matter.
We anticipate that new data from a variety of experimental and
observational fronts will soon bring us closer in unraveling the
particle nature of dark matter and the symmetry which protects its
lifetime.

\subsubsection*{Acknowledgements}

We thank Francesco D'Eramo, Andreas Hohenegger, Chris Jillings,
Alexander Kartavtsev, Rafael Lang, Maxim Pospelov, and Jesse Thaler
for helpful discussions and correspondence. Research at the Perimeter
Institute is supported in part by the Government of Canada through
NSERC and by the Province of Ontario through MEDT. Research at
Max-Planck-Institut f${\ddot{\it u}}$r Kernphysik is supported by
DFG-Sonderforschungsbereich Transregio~27.

\appendix*
\def\theequation{\thesection A.\arabic{equation}}
\setcounter{equation}{0}
\section{A~~~Boltzmann Equations}
\label{A}

The Boltzmann equation describing the evolution of the number density
$n_i$ is given by
\begin{equation}
 \frac{d n_i}{dt}+3 H n_i = C_i, 
\end{equation}
where $i=\eta,\,\chi$ and $ H(T) $ is the Hubble parameter accounting
for dilution of $n_i$ due to the cosmological expansion of the
Universe. The collision terms $C_i$ have the following contributions:
\begin{align}
 C_\eta &= C_{\eta\eta\rightarrow X_{SM}} +  C_{\eta\eta\rightarrow \chi\chi^*}
+ C_{\eta\chi\rightarrow \chi^*\chi^*} \nonumber \\
& +~C_{\eta\chi^*\rightarrow \chi\chi}
 +C_{\eta\rightarrow \chi\chi\chi}+C_{\eta\rightarrow \chi^*\chi^*\chi^*}~, 
\label{ceta}   
 \\[3pt]
C_\chi &= C_{\chi \chi^*\rightarrow X_{SM}} +  C_{\chi \chi   \rightarrow h \chi^*}
+ C_{\chi h   \rightarrow \chi^* \chi^*} 
+ C_{\chi \chi^*\rightarrow \eta\eta} \nonumber \\
& +~C_{\chi\eta\rightarrow \chi^* \chi^*} +C_{\chi\chi\rightarrow \eta\chi^*} + 
C_{\chi \chi \chi \rightarrow \eta}~. 
\label{cchi}
\end{align}

If the reheating temperature of the radiation dominated Universe was
high enough, all particles in consideration were once in thermal
equilibrium. Assuming Maxwell-Boltzmann statistics and instant
thermalization of the reaction products, $C_i$ can be written in
familiar, integrated form. Furthermore, $CP$ invariance relates
various collision terms so that we find for the individual
contributions of $C_{\eta}$:
\begin{eqnarray}
 C_{\eta\eta \rightarrow  X_{SM}} &=& -\langle \sigma v \rangle_{\eta\eta\rightarrow X_{SM}} 
\left[ n_\eta^2 -{(n_\eta^{{\rm eq}})}^2 \right], \nonumber \\
 C_{\eta\eta\rightarrow \chi\chi^*} &=& -\langle \sigma v \rangle_{\eta\eta\rightarrow \chi\chi^*}
 \left[ n_\eta^2 -\frac{n_\chi^2}{ {(n_\chi^{{\rm eq}})}^2 }{(n_\eta^{{\rm eq}})}^2\right],  \nonumber \\
 C_{\eta\chi \rightarrow \chi^*\chi^*} &=& -\langle \sigma v \rangle_{\eta\chi \rightarrow \chi^*\chi^*} 
\left[ n_\eta n_\chi -\frac{n_\chi^2}{ {n_\chi^{{\rm eq}}} }n_\eta^{\rm eq} \right], \nonumber \\
 C_{\eta\chi^* \rightarrow \chi\chi} &=&C_{\eta\chi \rightarrow \chi^*\chi^*}, \nonumber  \\
 C_{\eta \rightarrow \chi\chi\chi} &=& - \langle \Gamma_{\eta \rightarrow 3\chi} \rangle
\left[n_\eta -\frac{n_\chi^3}{{(n_\chi^{\rm eq})}^3}n_\eta^{\rm eq} \right], \nonumber \\
 C_{\eta \rightarrow \chi^*\chi^*\chi^*} &=& C_{\eta \rightarrow \chi\chi\chi} ,
\end{eqnarray}
while the terms  in $C_\chi$ in Eq.~(\ref{cchi}) are written as
\begin{eqnarray}
 C_{\chi \chi^*\rightarrow X_{SM}} &=& -\langle \sigma v \rangle_{\chi \chi^*\rightarrow X_{SM}}
 \left[ n_\chi^2 -{(n_\chi^{\rm eq})}^2 \right], \nonumber\\
C_{\chi \chi\rightarrow h \chi^*} &=& -\langle \sigma v \rangle_{\chi \chi\rightarrow h \chi^*}
 \left[ n_\chi^2 -n_\chi {n_\chi^{\rm eq}} \right], \nonumber \\
C_{\chi h\rightarrow \chi^* \chi^*} &=& -\frac{1}{2} C_{\chi \chi\rightarrow h \chi^*}, \nonumber\\
 C_{\chi \chi^*\rightarrow \eta\eta} &=& -\langle \sigma v \rangle_{\chi \chi^*\rightarrow \eta\eta}
 \left[ n_\chi^2 - \frac{n_\eta^2}{ {(n_\eta^{{\rm eq}})}^2 } {(n_\chi^{\rm eq})}^2 \right], \nonumber \\
 C_{\chi\chi\rightarrow \eta\chi^*}  &=& -\langle \sigma v \rangle_{\chi\chi\rightarrow \eta\chi^*} 
 \left[ n_\chi^2 - \frac{n_\eta}{n_\eta^{\rm eq}} n_\chi n_\chi^{\rm eq} \right], \nonumber \\
C_{\chi\eta\rightarrow \chi^* \chi^*} &=& -\frac{1}{2} C_{\chi\chi\rightarrow \eta\chi^*} , \nonumber \\
C_{\chi\chi\chi\rightarrow \eta} & = & -3 C_{\eta \rightarrow 3 \chi}  .  
\end{eqnarray}

The thermally averaged annihilation cross section for process
$a+b\rightarrow c+d$ is in general given by
\begin{eqnarray}
\langle \sigma v \rangle_{ab}& =& \frac{1}{8 m_a^2 m_b^2 T K_2(m_a/T)K_2(m_b/T)}  \nonumber \\
&\times & \int_{(m_a+m_b)^2}^{\infty} \!\! ds \, W_{ab} p_{ab}  K_1\left(\frac{\sqrt{s}}{T}\right),
\end{eqnarray}
where the quantity $W_{ab}$ is defined as
\begin{eqnarray}
 W_{ab} 
&=& \frac{1}{g_a g_b} \left( \frac{1}{8 \pi S_{cd}} \frac{\lambda^{1/2}(s,m_c^2, m_d^2)}{s}  \right)
\nonumber \\ 
&\times&
\sum_{spins} 
\int_{-1}^{1} \frac{d\!\cos{\theta}}{2} ~|{\cal M}_{ab \rightarrow cd}(\cos{\theta})|^2 .
\end{eqnarray}
In $W_{ab}$ the squared matrix element $|{\cal M}_{ab \rightarrow
  cd}|^2 $ is integrated over the c.m.~scattering angle
$\theta$. $K_i$ denotes the modified Bessel function of order $i$ and
$p_{ab}$ is the momentum of relative motion of $a$ and $b$.
Furthermore, $g_a$, $g_b$ are the number of internal degrees of
freedom, $S_{cd}$ is a symmetry factor accounting for identical final
states, and $\lambda(\hat x,\hat y,\hat z)\equiv\hat x^2 +\hat y^2
+\hat z^2 -2 \hat x \hat y - 2 \hat x \hat z - 2 \hat y \hat z$. For
the decay $a \rightarrow b + c+ \dots$, the thermal average reads
\begin{equation}
\langle \Gamma_{a \rightarrow b+c+\dots} \rangle = \frac{K_1(m_a/T)}{K_2(m_a/T)}\Gamma_{a \rightarrow b+c+\dots},
\end{equation}
where $\Gamma_{a \rightarrow b+c+\dots}$ is the standard decay rate
for a particle $a$ at rest.

\end{document}